# Fast electro-optic switching for coherent laser ranging and velocimetry

B. Haylock,[1] M.A. Baker,[2] T.M. Stace,[2,*] and M. Lobino[1,3,*]

[1]*Centre for Quantum Computation & Communication Technology and Centre for Quantum Dynamics, Griffith University, Brisbane QLD 4111, Australia*

[2]*ARC Centre of Excellence for Engineered Quantum Systems (EQUS), School of Mathematics and Physics, University of Queensland, Brisbane, 4072, Australia*

[3]*Queensland Micro- and Nanotechnology Centre, Griffith University, Brisbane, 4111, Australia*

The growth of 3D imaging across a range of sectors has driven a demand for high performance beam steering techniques. Fields as diverse as autonomous vehicles and medical imaging can benefit from a high speed, adaptable method of beam steering. We present a monolithic, sub-microsecond electro-optic switch as a solution satisfying the need for reliability, speed, dynamic addressability and compactness. Here we demonstrate a laboratory-scale, solid-state lidar pointing system, using the electro-optic switch to launch modulated coherent light into free space, and then to collect the reflected signal. We use coherent detection of the reflected light to simultaneously extract the range and axial velocity of targets at each of several electronically addressable output ports.

Optical scanners, capable of high-speed optical beam pointing are essential for many imaging techniques, including lidar and medical imaging applications. The first commercial demonstrations of multi-pixel lidar sensors relied on mechanical spinning mirrors, which are cumbersome and lack dynamic addressability. Similarly full-field optical coherence tomography relied on sample stage movement, or mechanical mirror steering to scan a sample. Recent advances have moved to simple integrated beam scanning techniques, including MEMS mirrors [1,2], optical phase arrays [3–5], and VCSELs [6,7]. Other major approaches include liquid crystal electro-optic scanners [8–10], electro-optic beam deflectors [11–13], and spectral scanning [14–16]. These new beam scanning techniques have allowed improved sensing performance by increasing the size and refresh rate of the generated point cloud. Most of these beam scanning technologies still limit the point cloud size and refresh rate due to speed limitations, with the notable exception of indium phosphide optical phase arrays, who have angle sweep rates of $> 10°/\mu s$, [17].

An alternative approach to spatial beam manipulation is to use a device with distinct separate spatial output modes to perform a discrete 'point-by-point' scan rather than a continuous sweep. A reconfigurable waveguide network can perform such a discrete scan. This approach ensures high speed, side-lobe-free, single mode, and single wavelength beam steering with the field of view and resolution set instead by the output optics. Such discrete scanning has previously been demonstrated with a silicon photonic integrated circuit, where the output channel is controlled thermally [18]. Here, we demonstrate a fibre-to-free-space multiplexing switch based on an integrated electro-optic device, which enables high speed, solid-state, single-mode output optical beam steering and light collection. We use this capability to demonstrate integrated laser ranging and velocimetry using coherent, modulated light and detection.

---

* Authors to whom correspondence should be addressed. Electronic mail: stace@physics.uq.edu.au. m.lobino@griffith.edu.au

The multiplexing switch is constructed from a network of directional couplers, and is fabricated by the reverse proton exchange technique in congruent lithium niobate [19,20]. The splitting ratio of each directional coupler is tunable between 0 and 100% by applying a voltage to electrodes patterned around two evanescently coupled waveguides. For this demonstration we create a switch with three output channels and total device loss of ~4dB. We characterize the frequency response of the one such electro-optic directional coupler by using it to amplitude modulate a laser beam. The modulated light is detected on a fast photodiode (PD), and the frequency response of the PD signal is shown in figure 1b, and demonstrates switching rates up to 300 MHz, limited by electrode design. With improved designs the response of such a system should reach the GHz range.

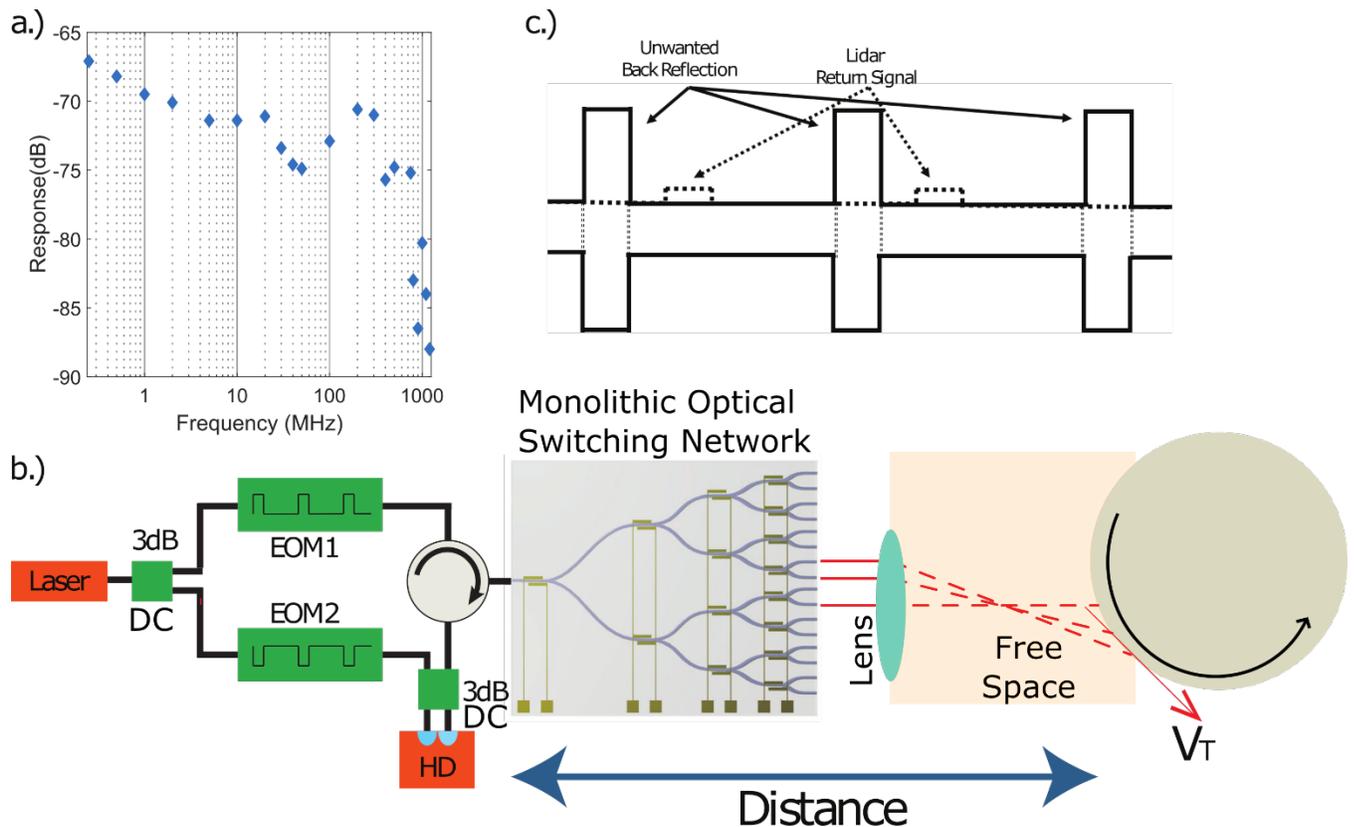

Figure 1 a) Sample frequency response of one electrode measured using a photodiode (DET08CFC, Thorlabs, 5GHz), with a sine wave injected from a waveform generator(E4432-B, Agilent, 250kHz-3GHz). The frequency response of the photodiode is not removed from the displayed signal. b) AMCW Lidar setup, full details in text. Here the first (signal) channel is shown at the top and the second (local oscillator) channel shown at the bottom. 3dB DC – 3dB directional coupler, EOM – electro-optic modulator, HD – homodyne detector, $V_T$ – tangential velocity of target. Light from the Switching Network is directed by a lens that maps the spatial location of the switch to a direction in free space. c) Timing diagram for signal modulation using EOM's, where the high level represents when the EOM transmits light

In what follows, we deploy this network switch for beam steering in an amplitude modulated continuous wave (AMCW) lidar system. Lidar systems are of great interest for automotive hazard detection and navigation, and a number of techniques have been implemented in practice. These include pulsed time of flight, frequency modulated continuous wave (FMCW), flash, and coherent flash protocols. Any system used for automotive purposes must support an encoding that can ignore interfering signals from other road users, such as the output of a chaotic laser for a 3D lidar system [24]. Pulsed systems can achieve this easily using random amplitude pulse encoding, however flash and FMCW systems are reliant on spectral separation, a scheme not suitable to produce millions of unique units. AMCW facilitates coherent time-of-flight plus Doppler analysis of the return signal, using eye-safe average and peak output powers. Our AMCW implementation allows for simultaneous measurement of range and velocity across discrete pixels in one dimension using only a single coherent detector. Additionally the switch architecture allows for temporal multiplexing of the detection electronics since all pixels are measured by a single detector. The full scheme is shown in Figure 1b.

A low noise CW laser at 1550nm (Koheras Boostik, NKT Photonics) is split by a 3dB coupler between two electro-optic modulators (EOMs). The EOMs are used to modulate the output light to provide time-of-flight sensitivity (see below), which is then passed through an optical circulator and into the custom-made monolithic 1-to-$N_{out}$ LiNbO$_3$ electro-optic switching network. For demonstration purposes, we choose $N_{out}$=3. The average power of each switch output is approximately 200 μW, limited by the maximum input power of the EOMs (100mW) and their transmission. To obtain higher output powers, and therefore better signal to noise ratios, a low noise erbium doped fiber amplifier should be used after the EOM in the signal arm, or a lower loss EOM. The output channels of the electro-optic switch are imaged through a lens onto a remote target. In this case, the test target was a spinning drum of radius $64 \pm 1mm$ with a diffuse surface placed approximately 4.7m away from the lens.

Depending on the transverse position of the switch output channel, the output of the switch is directed by the lens into rays in free space (see Fig. 1b), which intercept one or more target objects in the far-field. In our case, the target range is approximately 10 meters. In the current setup, the transverse separation between light from channels 1 and 2 is 9.4mm, and between channels 2 and 3 is 4.7mm at the target. These parameters are easily varied using different optics. After reflection from the object, the reflected light is collected at the same switch port from which it was launched. The collected light is recombined with the other arm of the 3 dB splitter, which acts as local oscillator, and is detected using a homodyne detector with bandwidth 100MHz [21]. Distance and velocity can be retrieved from the Fourier spectrum of the homodyne detector signal, which is captured at a sampling rate of 200MSPS using an oscilloscope.

The signal channel EOM is modulated to produce a square-wave, coherent pulse train on the output light, with repetition rate of 10MHz and pulse width of 10ns. Including intrinsic losses, the EOM transmission is -18.8dB. The second channel's EOM is modulated to produce the inverse pulse sequence of the first, with a delay suited so that back-reflection off the input of the electro-optic switch is not mixed with light from the other arm before being incident on the detectors as demonstrated in Figure 1c. The total loss of the second EOM is -6.1dB. This modulation of the reference signal reduces the detection of the back reflection from the electro-optic switch and allows an increase in the dynamic range of the lidar measurement. In future designs, the back reflection could be minimised by using a standard anti-reflection coating on the facets of the chip.

The monolithic electro-optic switching network is programmed to launch and collect light from each output channel in cyclic order, i.e. a measurement is taken from channel 1, then channel 2, and so on. A timing signal is sent from the switch control electronics to the oscilloscope to indicate which channel the received signal originates from.

Both range and velocity information can be extracted from the homodyne measurement of the collected light: time-of-flight ranging information is encoded in the time-delay of the pulse train relative to the launch time. From the relative phase of the signal reflected off the input of the electro-optic switch and the return signal, the time delay, and therefore the distance to target can be calculated. Axial velocity of the target is encoded in the Doppler frequency shift from the relative motion of the object, given a Doppler frequency shift $\Delta f$, the velocity of the target is given by $v = \lambda \Delta f = 1.55 \times 10^{-6} \Delta f \ ms^{-1}$. These quantities are extracted from the Fourier transform of the homodyne signal, examples of which are shown in Figure 2a).

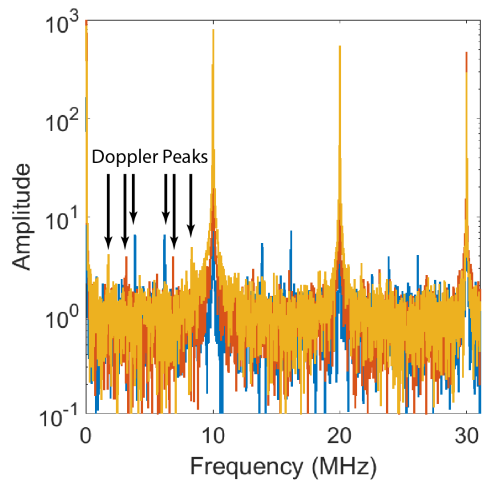

Figure 2 Sample return waveform from AMCW Lidar System. Each colour represents a different channel. Large peaks at multiples of 10MHz are caused by the reflected signal from the input of the optical switch. These are used as a timing reference. Other smaller peaks are the signal returned from the target that has been Doppler shifted.

The uncertainty in range and velocity is determined by the number of samples per channel and the signal to noise ratio. For these measurements the switching rate between different outputs is set to 10 kHz, with a duty cycle of 10% per channel, meaning 10000 samples are taken per channel per measurement. The frequency (velocity) resolution is calculated from the full width at half maximum of the lowest frequency Doppler peak, which was $\pm 33 kHz (\pm 25 mm/s)$. The position resolution is set by the sampling rate (200MSPS, $\pm 0.75 m$), however the random phase of the noise in the signal increases this uncertainty proportional to the signal to noise ratio (SNR). The accuracy can be improved by using a higher sampling rate, longer acquisition times or averaging over many acquisitions.

From the Doppler lidar measurements, using a known radius, the tangential velocity of the spinning target can be calculated as well as the angle-of-incidence for the light. The surface velocity of the target was set to $v_s = 15.3 \pm 0.1\ m/s$, and based on the angle of incidence, we calculate the axial target velocity $v_t = v_s \sin(\theta_{inc})$ to compare with the Doppler inferred velocity. These measured and computed quantities are summarized in Table 1, showing very good agreement between the measured and inferred velocities. For ranging, in our demonstration, the return signal was sufficiently weak that ranging data based on the phase shift was subject to large uncertainty. However, time-domain analysis of the return signal is capable of giving much improved ranging accuracy.

| Channel | Incidence Angle, $\theta_{inc}$ | SNR (dB) | Velocity ($m/s$) Doppler-measured, [Calculated, $v_t = v_s \sin(\theta_{inc})$] | Range (m) |
|---|---|---|---|---|
| 1 | 23° | 7.1 | 5.97±0.03, [5.98] | 4.5±0.6m |
| 2 | 19° | 4.3 | 4.91±0.03, [4.98] | 5.8±1.3m |
| 3 | 10° | 4.9 | 2.64±0.03, [2.66] | 4.5±1.1m |

Table 1 Summary of measurement results for three channel AMCW lidar. Range uncertainty limited by both sampling rate and SNR. The target was located a distance D=4.7m from the output.

Because of the low pulse repetition rate the unambiguous velocity measurement range is only $7.75 ms^{-1}$, and unambiguous range measurement interval is only 15m. This can be improved with higher pulse repetition rates and improved modulation schemes respectively. This beam steering method is also suitable for simultaneous range and velocity measurement using alternative lidar schemes.

An advantage of a discrete beam steering scheme such as the one presented here is the possibility for distributed sensing heads for lidar. As the outputs of the electro-optic switch may be easily recollected into fiber, the outputs can be easily routed to different places around a vehicle, with each fiber terminated by a taper or microlens. This means the 'sensor head' can be extremely small, unobtrusive and easily replaced. The TX/RX module, and the electro-optic switch may be integrated onto a daughter board connected to the central processing unit. This is ideally suited to harsh environment sensing, where any external sensor head may be easily damaged, and must be cheaply and easily replaceable. Furthermore the requirement of only one set of detection electronics for multiple sensors, can reduce the total sensor cost for automotive lidar.

## CONCLUSION

We have demonstrated the utility of high speed electro-optic switches for free-space beam scanning applications by creating a multi-pixel lidar system with a three channel switch. This scheme has several advantages including its speed, modularity, dynamic addressability, single-mode output, and compatibility with various coherent light modulation techniques. The approach we describe is scalable up to ~1000 output channels per chip using smaller footprint thin-film ridge waveguides in lithium niobate [22]. Additionally it could be considered to complement such a switching chip with acousto-optic deflectors on the waveguide outputs [23], to create a chip with 2D scanning, one direction discrete, and the other continuous.

## ACKNOWLEDGEMENTS

The authors thank Stefan Morley for electronics support. BH was supported by the Australian Government Research Training Program Scholarship. This research is financially supported by the Griffith University Research Infrastructure Programme, the Australian Research Council Centre of Excellence for Quantum Computation and Communication Technology (CQC2T, CE170100012), TMS and MAB were supported the Australian Research Council Centre of Excellence for Engineered Quantum Systems (EQUS, CE170100009), and ML was supported by the Australian Research Council Future Fellowship (FT180100055). This work was performed in part at the Queensland node of the Australian National Fabrication Facility, a company established under the National Collaborative Research Infrastructure Strategy to provide nano- and microfabrication facilities for Australia's researchers.

## REFERENCES

1. S. T. S. Holmström, U. Baran, and H. Urey, "MEMS laser scanners: A review," J. Microelectromechanical Syst. **23**, 259–275 (2014).
2. Blickfield GmbH, "Blickfield GmbH - Technology," https://www.blickfeld.com/tech-sol.
3. F. Xiao, W. Hu, and A. Xu, "Optical phased-array beam steering controlled by wavelength," Appl. Opt. **44**, 5429–5433 (2005).
4. K. Van Acoleyen, H. Rogier, and R. Baets, "Two-dimensional optical phased array antenna on silicon-on-insulator," Opt. Lett. **18**, 265–298 (2010).
5. J. Sun, E. Timurdogan, A. Yaacobi, E. S. Hosseini, and M. R. Watts, "Large-scale nanophotonic phased array," Nature **493**, 195–199 (2013).
6. N. W. Carlson, G. A. Evans, R. Amantea, S. L. Palfrey, J. M. Hammer, M. Lurie, L. A. Carr, F. Z. Hawrylo, E. A. James, C. J. Kaiser, J. B. Kirk, and W. F. Reichert, "Electronic beam steering in monolithic grating-surface-emitting diode laser arrays," Appl. Phys. Lett. **53**, 2275–2277 (1988).
7. K. Shimura, Z. Ho, M. Nakahama, X. Gu, A. Matsutani, and F. Koyama, "Non-mechanical beam scanner integrated VCSEL for solid state LiDAR," 2017 Conf. Lasers Electro-Optics Pacific Rim, CLEO-PR 2017 **2017-Janua**, 1–2 (2017).
8. C. Hu, J. R. Whinnery, and M. Nabil, "Optical Deflection in Thin-Film Nematic-Liquid-Crystal Waveguides," IEEE J. Quantum Electron. **QE-10**, 218–222 (1974).
9. S. Khan and N. Riza, "Demonstration of 3-dimensional wide angle laser beam scanner using liquid crystals," Opt. Express **12**, 868–882 (2004).
10. B. Luey, S. R. Davis, S. D. Rommel, D. Gann, J. Gamble, M. Ziemkiewicz, M. Anderson, and R. Paine, "A Lightweight, Cost-Efficient, Solid-State LiDAR System Utilizing Liquid Crystal Technology for Laser Beam Steering for Advanced Driver Assistance," 25th Int. Tech. Conf. Enhanc. Saf. Veh. Natl. Highw. Traffic Saf. Adm. 1–9 (2017).
11. R. A. Meyer, "Optical Beam Steering Using a Multichannel Lithium Tantalate Crystal," Appl. Opt. **11**, 613–616 (1972).
12. K. Nakamura, J. Miyazu, M. Sasaura, and K. Fujiura, "Wide-angle, low-voltage electro-optic beam deflection based on space-charge-controlled mode of electrical conduction in $KTa_{1-x}Nb_xO_3$," Appl. Phys. Lett. **89**, (2006).
13. D. B. Coyle, D. L. Rabine, D. Poulios, J. B. Blair, P. R. Stysley, R. Kay, G. Clarke, J. Bufton, and N. Goddard,

"Fiber Scanning Array for 3 Dimensional Topographic Imaging," **1**, 3–4 (2013).
14. X. Gu, T. Shimada, A. Matsutani, and F. Koyama, "Miniature nonmechanical beam deflector based on bragg reflector waveguide with a number of resolution points larger than 1000," IEEE Photonics J. **4**, 1712–1719 (2012).
15. A. Töws and A. Kurtz, "Investigations on the performance of lidar measurements with different pulse shapes using a multi-channel Doppler lidar system," in *19th Coherent Laser Radar Conference* (2018), pp. 1–5.
16. Baraja Pty Ltd, "Technology - Baraja," https://www.baraja.com/technology/.
17. Weihua Guo, P. R. A. Binetti, C. Althouse, M. L. Masanovic, H. P. M. M. Ambrosius, L. A. Johansson, and L. A. Coldren, "Two-Dimensional Optical Beam Steering With InP-Based Photonic Integrated Circuits," IEEE J. Sel. Top. Quantum Electron. **19**, 6100212–6100212 (2013).
18. A. Martin, J. Bourderionnet, L. Leviander, J. F. Parsons, M. Silver, and P. Feneyrou, "Coherent Lidar for 3D-imaging through obscurants," in *19th Coherent Laser Radar Conference* (2018), pp. 19–23.
19. J. L. Jackel and J. J. Johnson, "Reverse exchange method for burying proton exchanged waveguides," Electron. Lett. **27**, 1360 (1991).
20. F. Lenzini, S. Kasture, B. Haylock, and M. Lobino, "Anisotropic model for the fabrication of annealed and reverse proton exchanged waveguides in congruent lithium niobate," Opt. Express **23**, 1748 (2015).
21. R. Kumar, E. Barrios, A. MacRae, E. Cairns, E. H. Huntington, and A. I. Lvovsky, "Versatile wideband balanced detector for quantum optical homodyne tomography," Opt. Commun. **285**, 5259–5267 (2012).
22. A. Boes, B. Corcoran, L. Chang, J. Bowers, and A. Mitchell, "Status and Potential of Lithium Niobate on Insulator (LNOI) for Photonic Integrated Circuits," Laser Photon. Rev. 1700256 (2018).
23. D. E. Smalley, Q. Y. J. Smithwick, V. M. Bove, J. Barabas, and S. Jolly, "Anisotropic leaky-mode modulator for holographic video displays," Nature **498**, 313–317 (2013).